# On the Relation between Perfect Tunneling and Band Gaps for SNG Metamaterial Structures

Arif Shahriar [1], M.R.C. Mahdy[2], Jubayer Shawon[1], Golam Dastegir Al-Quaderi[3] and M. A. Matin[*1]

1 Department of EEE, Bangladesh University of Engineering & Technology, Dhaka, Bangladesh
2 Department of ECE, National University of Singapore, Singapore 117576, Singapore.
3 Department of Physics, University of Dhaka, Dhaka, Bangladesh.
*Corresponding author: M. A. Matin, email: *amatin@eee.buet.ac.bd*

In this article, we have proposed a compact classification of isotropic and homogeneous single negative (SNG) electromagnetic metamaterial based perfect tunneling unit cells. This unified classification has been made by means of the band gap theories and properties of the arrays made up of these unit cells. Based on their reported characteristics, we have proposed new structures that simultaneously show perfect tunneling band and complete band gap (CBG-omni directional stop band for both polarizations). Besides, we have identified a kind of perfect tunneling which can be considered as 'phase shifted perfect tunneling'. Several interesting and new phenomena like Complete Perfect Tunneling (CPT-omni directional perfect tunneling for both polarizations), Band Gap Shifting,CBG in Double Positive (DPS) material range, etc. have been reported with proper physical and mathematical explanations.

## Introduction:

Electromagnetic (EM) waves cannot propagate over several skin depths in materials exhibiting negative permittivity or negative permeability values (also known as single-negative (SNG) materials). EM waves that enter SNG materials exponentially decay along the direction of propagation and rapidly lose their energy. However, when SNG slabs are used in suitably designed multi-layer structures, they can be remarkably made completely transparent. It is also interesting that the combination of these unit cells into arrays can exhibit fully opposite behavior of transparency i.e. opaqueness or band gap behavior. Zhou et al. demonstrated a unit cell structure with high negative permittivity layer sandwiched between high positive permittivity medium which exhibits perfect tunneling [1]. Different perfect tunneling unit cells have also been demonstrated [2-4]. Alu-Engheta showed perfect tunneling with SNG bi-layers [5]. But the design proposed by Alu et al. works only at the sub-wavelength limit. Recently the idea of prefect tunneling using metamaterial has also been realized in semiconductor hetero-structure [6,7]using high potential and negative effective mass concepts. On the opposite side, to remove the problems or limitations of photonic crystals based photonic band gaps (i.e. Bragg gaps) Li et al. [8] and Jiang et al. [9] proposed non-Bragg gaps based on DNG and SNG metamaterials, respectively. Shadrivov et al. [10] demonstrated the idea of complete 3D band gap with only one dimensional periodic structure using metamaterials. But strangely no attempt has been made to relate all these independent works together. Although it seems that previous individual works are different, they may be related with each other. Moreover, no attempt has been taken to relate the perfect tunneling structures with band gap structures (neither in electromagnetics nor in semiconductor heterostructures). As a result, their engineering applications are very rare. Despite that metamaterial based perfect tunneling structures and band gaps have several attractive characteristics. Recently authors in Ref.[11] (one of the authors of this article) have found that metamaterial band gap inspired fibers with two different layers of unit cells can overcome the limitations of conventional Bragg fibers. Interestingly, we have also identified that the perfect tunneling reported in [2,3] should be considered as phase shifted perfect tunneling instead of pure perfect tunneling. Thus, it is imperative to answer the following notable questions:(i) Can we classify SNG metamaterial based perfect tunneling structures in a compact way so that we can understand their physical behavior and implement appropriate engineering applications? E. g. if unit cells made up of three or more layers to design array structures, we may get greater advantages in comparison with [11].(ii) More specifically, is there any relation between SNG metamaterial based perfect tunneling and their associated band gaps?(iii) Is it possible to achieve Perfect Tunneling Band (PTB), Complete Band Gap (CBG) etc. in a single structure? Such a structure can be very interesting for controlling the transmitted data according to the designer's will (especially for fiber clad, for fiber core and for broad band data transmission [11-13]).

In this article, we have investigated the answers of the above mentioned important questions. Here we have classified SNG metamaterial based perfect tunneling tri-layer unit cells by means of their band gap properties. Several interesting and new phenomena like Phase Shifted Tunneling, Complete Perfect Tunneling (CPT-omni directional perfect tunneling for both polarizations), Band Gap Shifting, CBG in Double Positive (DPS) materials etc. have been reported with physical and mathematical explanations. Our electromagnetic ideas can also be applied to semiconductor structures for tunneling and new type of artificial electronic band gaps.

**Theoretical Formulation:**

In Ref.[1], it has been shown that to achieve perfect tunneling, the reflection co-efficient, $r = -Q_{21}/Q_{22}$ should be zero provided there is no absorption present (where $Q(\omega,k)$ is a 2×2 transfer matrix that relates the forward backward electric field components in the incident medium to those of the transmitted medium [14]). But it does not give any clue to how this perfect tunneling is related with the band theory. To relate them and to investigate more physical insight, we can start considering the propagation along z-axis of the 1D photonic crystal (PC) [15], of a linearly polarized field of the form $\vec{E}_x(z,t) = E(z) \cdot \exp(-j\omega t)\hat{x}$. Starting from the Maxwell's equations:

$$\vec{\nabla} \times (\vec{\nabla} \times \vec{E}) + \frac{\mu(r)\epsilon(r)}{c^2} \cdot \frac{\partial^2 \vec{E}}{\partial t^2} - \frac{\vec{\nabla}\mu(r)}{\mu(r)} \times (\vec{\nabla} \times \vec{E}) = 0 \quad (1)$$

where $\epsilon(r)$ and $\mu(r)$ are the position dependent dielectric constant and magnetic permeability, respectively. Due to the periodicity of the 1D PC, $\epsilon(z+L) = \epsilon(z)$, $\mu(z+L) = \mu(z)$ where $L$ is the period, equation (1) can be reduced to:

$$\frac{d}{dz}\left(\frac{1}{n(z)Z(z)}\frac{dE(z)}{dz}\right) = -\frac{n(z)}{Z(z)} \cdot \frac{\omega^2}{c^2} E(z) \quad (2)$$

where $n(z) = \sqrt{\epsilon(z)} \cdot \sqrt{\mu(z)}$ and $Z(z) = \sqrt{\mu(z)}/\sqrt{\epsilon(z)}$. Considering the above form of the electric field $E_x(z,t)$, the solution of equation (2) for the electric field within each host material can be written as:

$$E(z) = E(z_0)\cos[k(z-z_0)] + \frac{1}{k}\sin[k(z-z_0)] \cdot \left(\frac{dE}{dz}\right)_{z=z_0} \quad (3)$$

where $k = 2\pi/\lambda = \omega/c \cdot |n|$ and $z_0$ denotes an arbitrary point in each layer. The above solution is valid only for stratified medium where $n(z)$ and $Z(z)$ are constants in a particular medium, but may vary from layer to layer. For a 1D PC composed of alternating layers of two or three or more materials, equation (2) must be solved by assuming the continuity of both $E(z)$ and $\frac{1}{n(z)Z(z)} \cdot \frac{dE(z)}{dz}$. This means that the two component function

$$\psi(z) = \begin{bmatrix} \psi_1 \\ \psi_2 \end{bmatrix} = \begin{bmatrix} E(z) \\ \frac{1}{nz} \cdot \frac{dE}{dz} \end{bmatrix} \quad (4)$$

is continuous through the PC structure. This condition can be written based on equation (3),

$$\psi(z) = M_i(z-z_0)\psi(z_0) \quad (5)$$

for each layer of the total structure. Here $i$ is the layer index, $z$ and $z_0$ are in the same layer and $M_i(z)$ can be recognized as the M matrix or transfer matrix defined by [15];

$$\begin{bmatrix} E_i \\ H_i \end{bmatrix} = M \begin{bmatrix} E'_{i+1} \\ H'_{i+1} \end{bmatrix} \quad (6)$$

Also from Eq (4), it is recognized that the term $\frac{1}{nz} \cdot \frac{dE}{dz} \propto H$ i.e. the magnetic field (Which can be shown very easily using Maxwell's equations). Hence for two-, three-or $m$-layer unit cells, according to [14], the total transfer matrix $T$ can be written as:

$$T = M_1 \cdot M_2 \cdots M_m = \begin{bmatrix} m_{11} & m_{12} \\ m_{21} & m_{22} \end{bmatrix} \quad (7)$$

where $M_i(\omega,k)$ is a 2×2 transfer matrix that relates the incident electric and magnetic field intensities and the transmitted intensities for the $i$-th layer ($M$ should be function of $k$). Again from [15], using Bloch's condition we have,

$$\psi(z+L) = \exp(iK_{BZ}L)\psi(z) \quad (8)$$

Hence, considering Eq (8) and the total array structure of $m$ unit cells of the form $(ABC)^n$ for a non-trivial solution of Eq (5) the eigen value equation gives:

$$\lambda^2 - (m_{11}+m_{22})\lambda + (m_{11}m_{22}-m_{12}m_{21}) = 0 \quad (9)$$

From conservation of energy it follows that $det(T) = 1 = (m_{11}m_{22}-m_{12}m_{21}) = \lambda_1\lambda_2$ and hence from Eq (8) we get:

$$\lambda_1 + \lambda_2 = \exp(jK_{BZ}L) + \exp(-jK_{BZ}L) = (m_{11}+m_{22})$$

$$\cos(K_{BZ}L) = \frac{1}{2}(m_{11}+m_{22}) \quad (10)$$

This is the so called band gap equation (although a different form has been given in [14]). Again from [16], using boundary conditions; it can be shown that:

$$\cos(KL) = \frac{1}{2}(q_{11}+q_{22}) \quad (11)$$

But we have not yet been able to relate the elements of Q matrix ($q_{21}$, $q_{11}$ etc.) with the elements of $T$ matrix. ($m_{21}$, $m_{11}$ etc.) So, going back to the basic definitions and we can

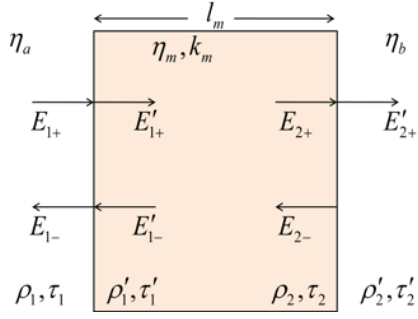

FIG-1. Structure of magneto dielectric slab with associated parameters for the detemination of the relation between the elements of $Q\&T$ matrices (See ref [15], too).

write from [15] (also see Fig. 1):

$$\begin{bmatrix} E_{1+} \\ E_{1-} \end{bmatrix} = Q \begin{bmatrix} E'_{2+} \\ E'_{2-} \end{bmatrix} \quad (12)$$

where

$$Q = Q_1 \cdot Q_2 \cdots Q_m = \begin{bmatrix} q_{11} & q_{12} \\ q_{21} & q_{22} \end{bmatrix} \quad (13)$$

We know the forward and backward waves are related $E_+(z) = \frac{1}{2}[E(z) + \eta(z)H(z)]$; $E_-(z) = \frac{1}{2}[E(z) - \eta(z)H(z)]$
So using the relation:

$$\begin{bmatrix} E_{1+} \\ E_{1-} \end{bmatrix} = \frac{1}{2}\begin{bmatrix} E_1 + \eta_a H_1 \\ E_2 - \eta_b H_1 \end{bmatrix} \text{ and } \begin{bmatrix} E'_1 \\ H'_2 \end{bmatrix} = \begin{bmatrix} 1 & 1 \\ \frac{1}{\eta_a} & -\frac{1}{\eta_b} \end{bmatrix}\begin{bmatrix} E'_{2+} \\ H'_{2-} \end{bmatrix}$$

Where $\eta_a$ and $\eta_b$ are the characteristic impedances of incident and transmitted media respectively. Now, we can write:

$$\begin{bmatrix} E_{1+} \\ E_{1-} \end{bmatrix} = \begin{bmatrix} \frac{1}{2}\{(m_{11} + \frac{\eta_a}{\eta_b}m_{22}) + (\eta_a m_{21} + \frac{1}{\eta_b}m_{12})\} & \frac{1}{2}\{(m_{11} - \frac{\eta_a}{\eta_b}m_{22}) + (\eta_a m_{21} - \frac{1}{\eta_b}m_{12})\} \\ \frac{1}{2}\{(m_{11} - \frac{\eta_a}{\eta_b}m_{22}) - (\eta_a m_{21} - \frac{1}{\eta_b}m_{12})\} & \frac{1}{2}\{(m_{11} + \frac{\eta_a}{\eta_b}m_{22}) - (\eta_a m_{21} + \frac{1}{\eta_b}m_{12})\} \end{bmatrix}\begin{bmatrix} E'_{2+} \\ E'_{2-} \end{bmatrix} \quad (14)$$

From (12) & (14), we can write:

$$\begin{bmatrix} q_{11} & q_{12} \\ q_{21} & q_{22} \end{bmatrix} = \begin{bmatrix} \frac{1}{2}\{(m_{11} + \frac{\eta_a}{\eta_b}m_{22}) + (\eta_a m_{21} + \frac{1}{\eta_b}m_{12})\} & \frac{1}{2}\{(m_{11} - \frac{\eta_a}{\eta_b}m_{22}) + (\eta_a m_{21} - \frac{1}{\eta_b}m_{12})\} \\ \frac{1}{2}\{(m_{11} - \frac{\eta_a}{\eta_b}m_{22}) - (\eta_a m_{21} - \frac{1}{\eta_b}m_{12})\} & \frac{1}{2}\{(m_{11} + \frac{\eta_a}{\eta_b}m_{22}) - (\eta_a m_{21} + \frac{1}{\eta_b}m_{12})\} \end{bmatrix} \quad (15)$$

Equation (15) is the most important equation which relates the elements of $Q\&T$ matrices. Based on the general condition of the perfect tunneling $Q_{21}(\omega,k) = -q_{21}(\omega,k) = 0$, as stated by in[1], we are going to classify SNG metamaterial based PT unit cells into different categories. But we need not consider TE & TM modes seperately since for normal incidence they are equivalent. As a result we are considering the relation between $q_{21}$ and $m_{ij}$ from Eq (15):

$$q_{21} = \frac{1}{2}(m_{11} - \frac{\eta_a}{\eta_b}m_{22}) - \frac{1}{2}(\eta_a m_{21} - \frac{1}{\eta_b}m_{12}) \quad (16)$$

where, $\eta_{a,b}$ are the characteristic impedances of the initial and final medium. The formula for inclined incidence is:

$$q_{21} = \frac{1}{2}(m_{11} - \frac{\eta_{aT}}{\eta_{bT}}m_{22}) - \frac{1}{2}(\eta_{aT} m_{21} - \frac{1}{\eta_{bT}}m_{12}) \quad (17)$$

where the transverse impedance $\eta_T$ stands for either $\eta_{TM}$ or $\eta_{TE}$ given by:

| Polarization | $\eta_T$ |
|---|---|
| $\eta_T$ (TE) | $\eta/\cos\theta$ |
| $\eta_T$ (TM) | $\eta\cos\theta$ |

Above discussion gives a single condition in terms of elements of $q$ (i.e $q_{21} = 0$) which is actually equivalent to different possibilities in terms of $m_{ij}$. We have used this very fact for the classification of different unit cell combinations of SNG metamaterials. It is the most notable point that throughout the article we have imposed primary conditions based on only equation (16) (i.e. for normal incidence) and considering $\eta_a = \eta_b$. But these primary conditions are strong enough to handle the situations of inclined incidences as well. On the other hand, equation (10) relates the band gap with perfect tunneling condition (Eq (16)) based on the simple relation: det($T$) =1. Considering these simple relations, we are goingto propose structures with simultaneous complete perfecttunneling band with complete band gaps just by using 1D layered structure.

Depending on Eq (16) and (10) our classification is as follows:

*Type-A1 and Type-A2 unit cells:*
Considering the 1st case of the table-1 with the condition: $\sum \epsilon_i d_i = \sum \mu_i d_i = \sum d_i/\epsilon_i = \sum d_i/\mu_i = 0$ &for electrically

| | Main & Primary condition of PT (always satisfied at $f=f_T$): $q_{21}(f_T)=0$ [Eq (16)] Band Gap condition: $\cos(K_{BZ}L) = (1/2)(m_{11}(f_T) + m_{22}(f_T))$; Connected condition: $\det(T) =1$ | | | | | | | | |
|---|---|---|---|---|---|---|---|---|---|
| Type | Minimum Condition to be satisfied at normal incidence only | $T=1$ for Both Polarizations [unitcell] | $T=1$ for all angle (0 to 90 degrees) [unit cell] | Perfect tunneling freq. [unit cell]=Band edge freq. of stop band [array] | Perfect tunnel-ing band (array) | Complete Band gap (array) | Band gap shifting (by adding matched layer with unit cell then arraying) | Examples of similar perfect tunneling unit cells | Comment |
| | **Specific condition:** $m_{12}=0$ **&** $m_{21}=0$ **(automatically satisfied for A1 & A2 at $f_T$)** | | | | | | | | |
| A1 | $\bar{\epsilon} = \bar{\mu} =$ $\overline{1/\epsilon} =$ $\overline{1/\mu} = 0$ (Electrically Thin layer) | Yes | Yes | Yes($K_{BZ}$ =0 )[But zero width BG; no real BG] Shifting of PT | Yes (must be >AB) [*CPT for *CMA 1] | Yes (always because of $\eta_1 = \eta_2$ = ……) | – | 1.Fig-5,11 (only AB case) in [5] 2.Fig2; case-1; in [4] | 1.Unit cell:*CPTB 2.Array: Simultaneous *PTB & *CBG is possible for > AB ( Best case: *CMA1) |
| A2 | $\bar{\epsilon} = \bar{\mu} = 0$ $\overline{1/\epsilon} \neq 0$ $\overline{1/\mu} \neq 0$ (Electrically Thin layer) | Yes | No | Yes($K_{Bz}$=0) [But zero width BG; no real BG] shifting of PT | Yes( must be >AB) | Yes (al-ways because of $\eta_1 = \eta_2$ = ……) | – | 1.Fig3 (only AB case); in [5] 2.Fig2; case-2,3; in [4] | 1.Unit cell: No CPTB 2.Array: very small PTB but good CBG is possible for > AB |
| B | $\bar{\epsilon} \neq 0,$ $\bar{\mu} \neq 0$ Electrically Thick&Thin layer($m_{12} = m_{21}=0$ ; &$m_{11} = m_{22}$) | Yes | Yes | Yes ($K_{Bz}$ =0 OR $K_{Bz}= \pi/L$) No Shifting of target PT frequency | No | For Specific cases | No | 1.[1] | 1.Unit cell:*CPT 2.Array: No PTB but CBG is possible for forced conditions > AB. |
| | **Specific condition:** $m_{12} \neq 0$ **&** $m_{21} \neq 0$ ; $q_{21}$=0[ **Phase Shifted Perfect Tunneling**] | | | | | | | | |
| C | $\bar{\epsilon} \neq 0,$ $\bar{\mu} \neq 0$ (Electrica-lly Thick&Thin layer) | No | No | No ($K_{Bz} \neq 0$ but must be real) No Shifting of target PT frequency | Yes | For Specific cases | Yes | 1.[2] 2.[3] 3.Fig-10,12,9(case-1,2&3);in [4] | 1.Unit cell: No CPT 2.Array: Simultaneous PTB & CBG is possible for forced conditions > AB |

**TABLE-1**

*CPTB: Complete Perfect Tunneling Band  *PTB: Perfect Tunneling Band  *CMA1: Conjugate Matched A1  *CBG: Complete Band gap
*CPT: Complete Perfect Tunneling at specific frequency

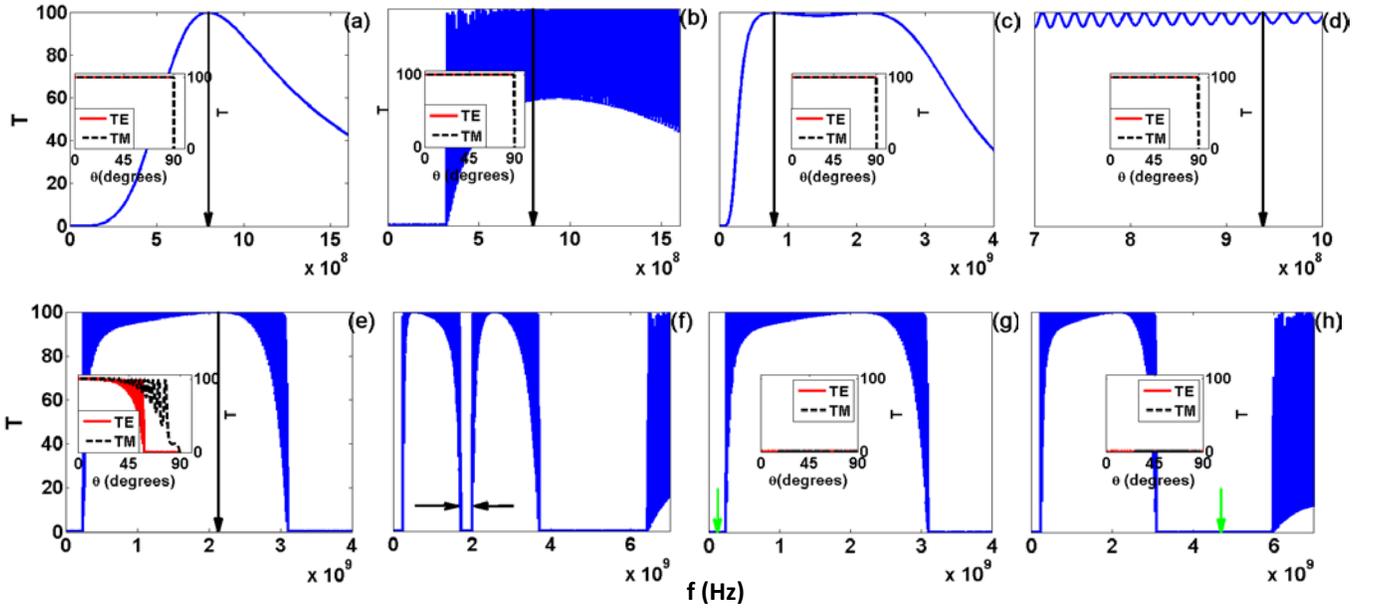

FIG-2.A1 type conjugated matched AB type unit cell supports CPT at target frequency, $f_0$ -2(a). But the array (100 unit cells) does not support PTB, 2(b). [Parameters: at target frequency, $f_0 = 0.796$ GHz, $\varepsilon_1 = -3$, $\mu_1 = 6$; $\varepsilon_2 = 3$ $\mu_2 = -6$; $d_1 = d_2 = 10$mm. Drude dispersive model has been used. Insets of 2(a) & 2(b) are drawn at the target frequency of $f_0 = 0.796.796$ GHz with angle of incidence, $\theta = 0$ degrees.] A1 type conjugated matched ABC unit cell supports CPTB -2(c). Althoug, the array(100 unit cells) supports CPT only at $f_0$, 2(d); it is capable of showing PTB, 2(e) [Prameters: at target frequency, $f_0 = 0.796$ GHz, $\varepsilon_1 = -2$ $\mu 1 = 4$,$\varepsilon_2 = 2$ $\mu_2 = -4$, $\varepsilon_3 = 2$ $\mu_3 = -4$; $d_1 = d_3 = 5$mm, $d_2 = 10$mm. (Insets of 2(c) & 2(d) are drawn at $f_0 = 0.796$ GHz and that of 2(e) is at 2.127 GHz.)

For oblique incidence a new BG opens up in between 1.7 GHz & 2 GHz, as shown in 2(f) (at $\theta = 45$ degrees) which is due to the apparent violation of Snell's law.

A1 type ABC unit cell array (100 unit cells support CBG. Here, 2(g) refers to an SNG gap and 2(h) refers to a DPS gap. [Parameters: same as Fig 2(c) & 2(d). Insets of 2(g) & 2(h) are drawn at the frequencies of 0.125GHz and 4.6GHz respectively]. Green line will indicate the mid gap frequency throughout the whole article (Black lines are PT frequency).

thin layer $(d << \lambda)$, we shall discuss the interesting properties of these unit cells (named type-A1) and the array made up of such unit cells. It is important to note that the condition : $m_{12} = 0$, $m_{21} = 0$; $m_{11} = m_{22}$ (see Eq (19) in [4]) are automatically satisfied. Also the impedance matching condition $\eta_1 = \eta_2 = .....$ is automatically satisfied at the PT frequency. Also the equal optical path condition[17] i.e. $k_1d_1 = k_2d_2 = .....$ is not a generalized necessary codition for PT. To get the conditions stated in Table-1 for Type-A1, we referto the idea of Spatial Averaged Single Negative (SASN) BG [18] (a special type of Zero Effective Phase(ZEP) band Gaps [9]). At subwavelength limit, if a unit cell is made up of $m$ layers and we make their array, it can be shown that $m_{11} = m_{22} = 1$ (see the $T$ matrix in [4]& [19]) . According to Eq (10), it turns into the condition: $\cos(K_{BZ}L) = 1$ or the band edge condition and the SASN band edge condition can be written in a very simple way :

$$1 - \frac{(\alpha_1 d_1)^2}{2} - \frac{(\alpha_2 d_2)^2}{2} - \frac{(\alpha_3 d_3)^2}{2} - \frac{1}{2}\left(\frac{F_2}{F_1} + \frac{F_1}{F_2}\right)\alpha_1 d_1 \alpha_2 d_2 - $$
$$\frac{1}{2}\left(\frac{F_3}{F_2} + \frac{F_2}{F_3}\right)\alpha_3 d_3 \alpha_2 d_2 - \frac{1}{2}\left(\frac{F_3}{F_1} + \frac{F_1}{F_3}\right)\alpha_1 d_1 \alpha_3 d_3 = \cos(K_{BZ}L) = 1 \quad (18)$$

Hence at band edge for TE mode:

$$\sum \epsilon_i d_i - \frac{ck_\parallel}{\omega}(\sum d_i/\mu_i) = 0 \text{ or } \sum \mu_i d_i = 0 \quad (19)$$

And for the TM mode:

$$\sum \epsilon_i d_i = 0 \text{ or } \sum \mu_i d_i - \frac{ck_\parallel}{\omega}(\sum d_i/\epsilon_i) = 0 \quad (20)$$

where $F_i$ (TM) $= \eta_{Ti}$ (TM), $F_i$ (TE) $= (1/\eta_{Ti}$ (TE)) and $\alpha_i = (\omega/c)(\sqrt{\varepsilon}\sqrt{\mu})\cos\theta$. A very nice application and extension criteria of such BG has been discussed in [11] & [19] respectively. Bandwidth of the SASN gap will be zero if we choose the above two conditions (19) & (20) for $f_1=f_2$ [19]. By combining both conditions together for any angle of incidence to achieve transparency at frequency $f_1=f_2=f$ and also for both TE & TM polarization the ultimate condition turns into $\sum \epsilon_i d_i = \sum \mu_i d_i = \sum d_i/\epsilon_i = \sum d_i/\mu_i = 0$ (applicable only for subwavelength case). This specific condition has been shown as the 'Alu-Engheta condition' for simplest AB layer unit cell in [5,17]. As $f_1 = f_2$, no real BG opens up and it has been described as the zero-width band gap [17]. This has an effective phase delay point i.e transparent media with zero phase lag and no real BG i.e. zero-width BG. However, we shall show that if the unit cells are made up of more than two layers (>AB), several unusual phenomena occur. These unusual phenomena are not possible with the AB unit cells.In Fig.2(a) & (b), we have shown that the transmittivity, $T = 1$ for both

polarizations(for the unit cell)at any angle of incidence at the target frequency $f_0$. But when the unit cell is made up of more than two layers (i.e. ABC structure), only then for a range of frequencies, nearly perfect tunneling occurs which we can call as a Perfect Tunneling Band (PTB). This is a new observation (Fig-2 (e)). Besides, when we make an array with the unit cells of ABC, the PTB still exists (although in the DPS region Fig. 2(h))which is not possible with two layers A1 type unit cell. However, Fig-2(c) & (e) clearly tell us thattype-A1 unit cells (> AB) can produce sufficientlywide PTB (even CPT for unit cell ) which could be very useful for fiber cores and sensors [12].

To understand how this PT is occuring in the DPS region even with electrically thick layers we calcute the phase of the transmitted electric field. We find that this is not pure PT rather it should be considered as Phase Shifted Perfect Tunneling (PSPT). Interestingly the same kind of phase shifted PT has been reported in a very complex way using metamaterials in Ref.[2,3]. However, in [2,3] this phenomenon – 'phase shifted PT' has not been explicitly discussed. In such PSPT cases, the transmitted wave gets an extra phase although the magnitude of the fields remain same (same power). Although for PSPT $q_{21} = 0$, it occurs only when $T$ matrix (Eq (7)) is not an identity matrix. In such cases, the PT frequency is found at the pass band for the arrays of the unit cell because $m_{12}$ & $m_{21} \neq 0$ for such cases ($m_{11}$ & $m_{22} \neq 1$). We have discussed in detail later about such case in Type-C unit cell description. Mathematically we have :

$$\begin{pmatrix} E' \\ H' \end{pmatrix} = T^{-1} \begin{pmatrix} E \\ H \end{pmatrix} = \begin{pmatrix} m_{11} & m_{12} \\ m_{21} & m_{22} \end{pmatrix}^{-1} \begin{pmatrix} E \\ H \end{pmatrix}$$

The transmittance is: $t = \dfrac{P_{Transmitted}}{P_{incident}} = \dfrac{|E'|^2}{|E|^2}$ (21)

If we consider $|t|^2 = 1$ then, $|E'| = |E|$. So, the only possibility is:

$$E' = E e^{i\beta} \quad (22)$$

where $\beta$ is the additional phase shift which does not occur in pure PT cases (when $T$ is an identity matrix).

As $m_{11}$ & $m_{22}$ are real but $m_{12}$ & $m_{21}$ are imaginary whenall the layers are DPS, from Eq (21) & (22), we can write at $T=1$:

$$\tilde{m}_{11} E + \tilde{m}_{12} H = E \cos\beta + iE \sin\beta \quad (23)$$

Implying that-

$$\tilde{m}_{11} = \cos\beta \text{ and } \tilde{m}_{12} = i\frac{E}{H}\sin\beta = i\eta_{in}\sin\beta$$

where $\eta_{in}$ is the impedance of the incident medium and $\tilde{m}_{ij}$ are the elements of the inverse matrix $T^{-1}$. Therefore-

$$\beta = \tan^{-1}\left(\frac{1}{\eta_{in} i}\frac{\tilde{m}_{12}}{\tilde{m}_{11}}\right) \quad (24)$$

Similarly, considering magnetic field; it can be shown that:

$$\alpha = \tan^{-1}\left(\frac{1}{\eta_{in} i}\frac{\tilde{m}_{21}}{\tilde{m}_{22}}\right) \quad (25)$$

The equations (24) & (25) have been satisfied at the DPS region with $m_{12} \neq 0$ and $m_{21} \neq 0$ in the perfect tunneling frequency ranges in Fig. 2(e). Field distributions in Fig. 3(e) & 3(f) clearly tells that decoupling mode theory [17] is

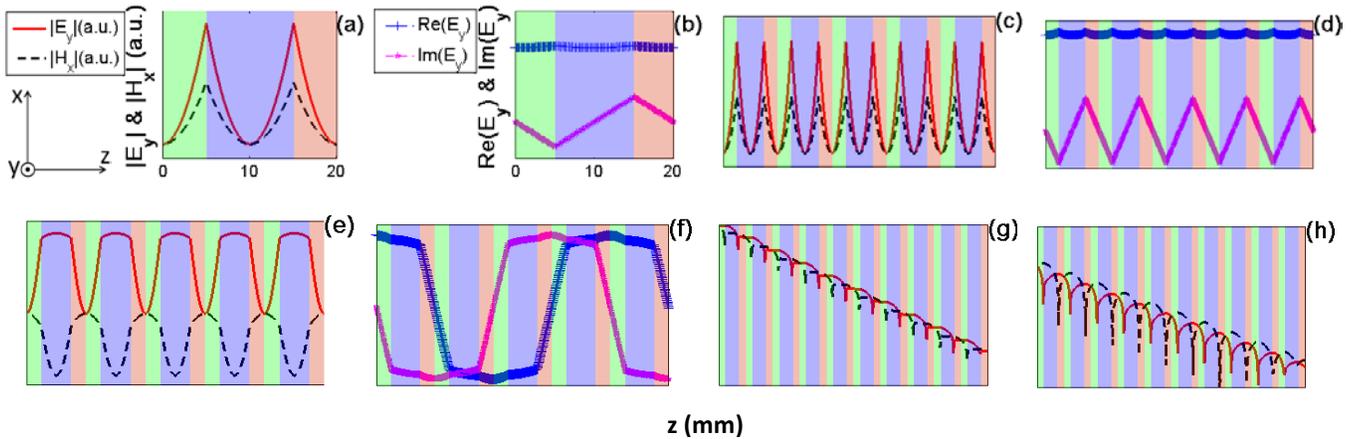

FIG-3. The $E$ and $H$ intensities as a function of position for A1 type ABC unit cell at $f_0$= 0.796GHz is shown in 3(a). 3(b) shows the real & the imaginary part of the $E$ field for the same configuration as 3(a). 3(c) & 3(d) are drawn for 6-unit cell array, for the same data. 3(e) & 3(f) are drawn for 6-unit cell array at 2.15GHz where PT of has been achieved for the array (at pass band), 3(g) is drawn for SNG BG (at mid frequency 0.125GHz in SNG stop band with 10 unit cell array in log scale). 3(h) is at mid 4.6GHz of DPS BG for 10 unit cell array in log scale.

not applicable for the PSPT due to strong reflection ($K_{Bz} \neq 0$ or $\pi/L$). In terms of the theory of decoupled modes, the target PT frequency $f_0$ at the SNG region is nothing but the frequency of the zero width bandgap [17]. The BG will open if and only if the Alu-Engheta condition [5,17] for $\epsilon$ is met for $f_1$ but for $\mu$ it is met at a different frequency $f_2$[19]. From the field distribution (Fig-3 (a)), it is clearly found that interaction of pure evanscent or decaying modes are responsible for this kind of phenomenon[20]. Actually these are the band edge modes (after arraying)as reported in [20] to explain the Type-II gap ( SASN BG [18]). Hence, our proposed condition can be treated as generalized Alu-Engheta condition for ENG|MNG $m$ layers to form unit cells for getting perfect tunneling over a wide frequency range i.e. a PTB(CPT at the target frequency). But the most noticable fact is that after making an array with the ABC unit cells, although we get CPT at the target frequency, PTB like that in the DPS region can not be found in SNG frequency range (Fig. 2(d)). The transmittivity is an oscillatory function of frequency near the target frequency. So, the statement reported in [17] : " ….. which results $K_{Bz}$ = 0.The latter quantity assures that the wave is restored after passing through the crystal" is true only for the target frequency. As $K_{Bz} = 0$ at that frequency,wave will be restored even after arraying the unit cells. Hence the array structure will achieve perfect tunneling at the zero band width frequency (Fig. 2 (b) & (d)). But there is still another way to achieve CPT at the target frequency and a PTB with the array. This condition can be called conjugate matched A1 (CMA1) condition:

$|\varepsilon_1| = \mu_1 = \varepsilon_2 = |\mu_2| = |\varepsilon_3| = \mu_3; d_1 = d_3 \& d_2 = 2d_1$ (ENG|MNG|ENG).For other frequencies of the PTB near the target frequency, the additional phase gain (Eq (24))

achieved by the fields are extremely small and $K_{Bz}$ remains almost zero at those frequencies. Hence the wave is restored almost at the pure from after passing through the array (Fig. 4(a)). This is a superior performance possible only with the CMA1 structures.

A completely different and quite strange property is the band gap property or opaqueness of such unit cells is achieved by putting them into arrays. Although for some specific frequency range they behave as perfect tunneling unit cells, we notice some very unusual and interesting phenomena in their BG properties for other frequency ranges. For example, in the DPS frequency range we are getting CBG which is a new observation (Fig.2(h)). For oblique incidence another BG opens up at frequencies near but greater than the plasma frequencies of the SNG metamaterials (Fig. 2(f)). At these frequencies the BG is not the conventional Bragg gap rather this gap is due to the fact that the angle of refraction in the layer is imaginary as can be seen from Snell's law [20]. This is like an aparent violation of the Snell's law. However we find that at some particular frequency the BG in the DPS region is omni-directional (inset of Fig .2(h)). This is a new observation that in the DPS region for some frequency range we get omni-directional Bragg gap.

For BG due to aparent violation of Snell's law it is necessary to use a third layer since the incident waves can not couple to Brewstar window and hence it can show the CBG performance even with the propagating modes (similar but not identical to the case of DNG based CBG reported in [10]).On the other hand, the BG originates in the SNG range (Fig 2(g)) is due to the pseudo-propagating modes. It also shows CBG property(Fig 2(g)) due to $F_1 = F_2 = F_3$ at both polarizations. According to the field

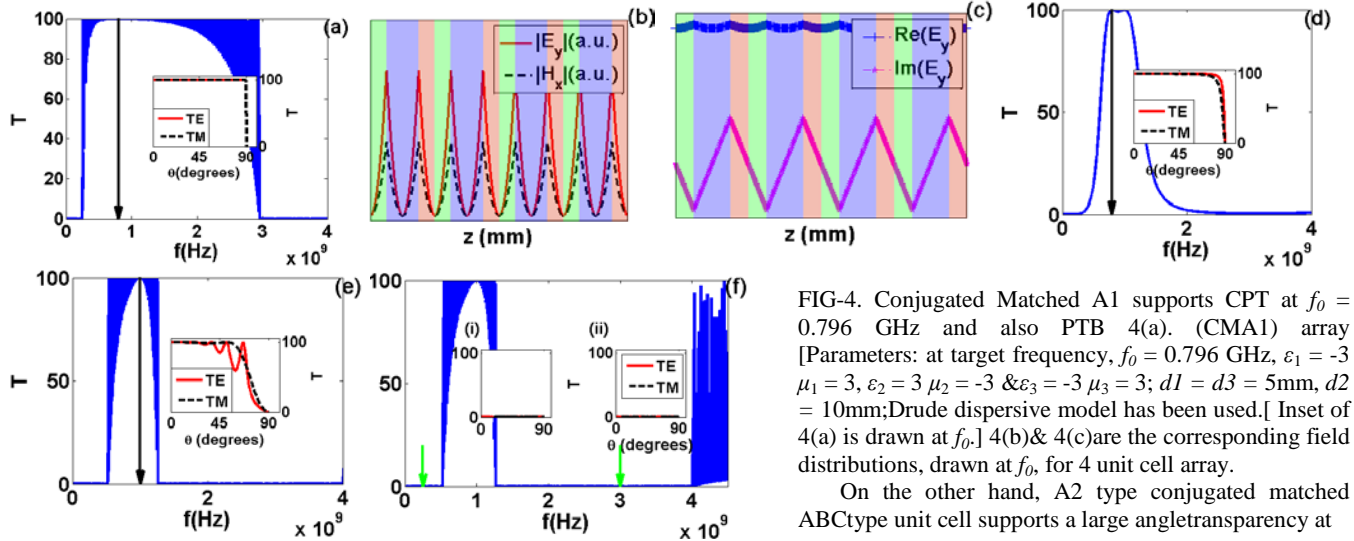

FIG-4. Conjugated Matched A1 supports CPT at $f_0$ = 0.796 GHz and also PTB 4(a). (CMA1) array [Parameters: at target frequency, $f_0$ = 0.796 GHz, $\varepsilon_1$ = -3 $\mu_1$ = 3, $\varepsilon_2$ = 3 $\mu_2$ = -3 & $\varepsilon_3$ = -3 $\mu_3$ = 3; $d1$ = $d3$ = 5mm, $d2$ = 10mm;Drude dispersive model has been used.[ Inset of 4(a) is drawn at $f_0$.] 4(b)& 4(c)are the corresponding field distributions, drawn at $f_0$, for 4 unit cell array.

On the other hand, A2 type conjugated matched ABCtype unit cell supports a large angletransparency at the target frequency $f_0$ - 4(d).The array(100 unit cells) supports PTB, 4(e) [Parameters: at target frequency, $f_0$= 0.796 GHz, $\varepsilon_1$ = -6 $\mu_1$ = 12, $\varepsilon_2$ = 3 $\mu_2$ = -6, $\varepsilon_3$ = -6 $\mu_3$ = 12; $d_1$ = $d_3$ = 5mm, $d_2$ = 20mm;Drude dispersive model has been used. Insets of 4(d)& (e) are drawn at the frequencies of 0.796 GHz &0.9983GHz respectively].A2 type ABC cell array of 100 unit cells support CBG, 4(f). Here, 4(f) inset-(i) refers to an SNG gap and 4(f) inset (ii) refers to a DPS gap [parameters: same as Fig 4(d) & 4(e). Insets of 4(f), (i) & (ii) are drawn at the frequencies of 0.25GHz and 3.0GHz respectively].

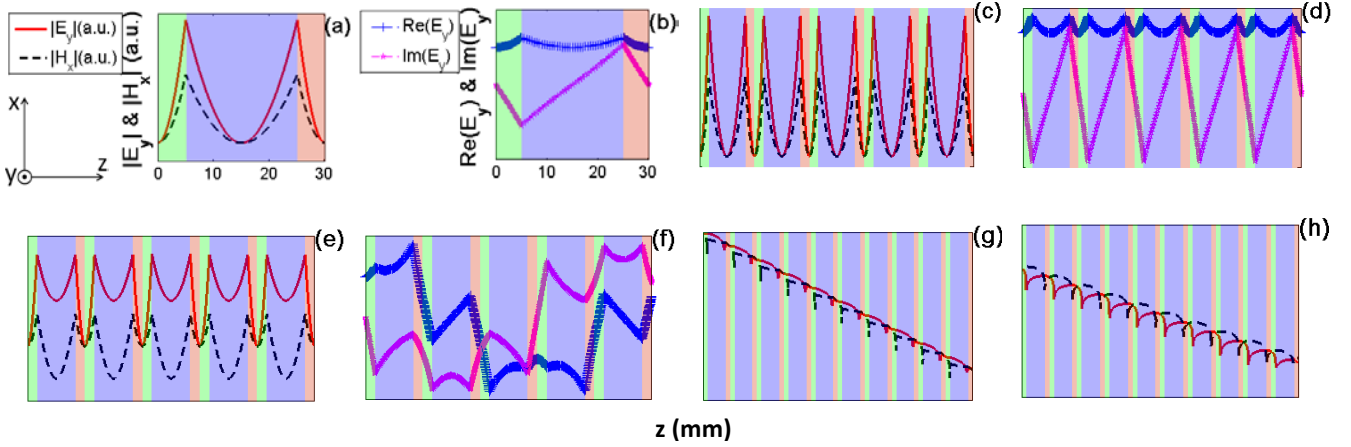

FIG. 5. The $E$ and $H$ intensities as a function of position for A2 type ABC unit cell at $f_0 = 0.796$GHz is shown in 5(a). 5(b) shows the corresponding real and imaginary parts of $E$ field with respect to the position. 5(c) & 5(d) are drawn for 5-unit cell array at $f_0$. 5(e) & 5(f) are drawn for 5-unit cell array at 0.9983GHz(at DPS pass band). 5(g) is drawn for SNG BG (at mid 0.25GHz in stop band with 10 unit cell array in log scale). 5(h) is at mid 3GHz of the DPS gap for 10 unit cell array in log scale.

distribution of Fig-3(g) , this SNG gap behaves like type-I BG reported in [20] which is different than SASN [18] or ZEP [9] BG. Both of these band gaps (Fig. 2(g) & 2(h)) are originated from type-A1 unit cells and are independent of polarizations and the angle of incidence.

Type-A1 unit cells (> AB) individually show superior perfect tunneling performance. However, if we introduce small amount of loss in the metamaterials the CPT will be destroyed. But the PTB will sustain even in the presence of loss. Their array is also superior due to simultaneous PTB (although shifted from target frequency and an additional phase is gained by the fields)& CBG. Most superior behavior can be expected from CMA1 unit cells and arrays. CMA1 arrays will show simultanous PTB with CPT at target frequency as well as CBG. Such designs could be very useful in the near future. Because of their coupling to Surface Plasmons (SP) and conventional modes in superfast metamaterial inspired fibers [12].

Consider the 2$^{nd}$ case of the Table-1 (type A2), which satisfies almost the same condition $\sum \epsilon_i d_i = \sum \mu_i d_i = 0$ & is electrically thin ($d \ll \lambda$ like A1 type ) but $\sum d_i/\epsilon_i \neq 0$ & $\sum d_i/\mu_i \neq 0$. Again note that the conditions: $m_{12} = 0$, $m_{21} = 0$; $m_{11} = m_{22} = 1$ (see Eq (19) in [4]) are automatically satisfied and $\eta_1 = \eta_2 = \ldots$ will be satisfied at the PT frequency. However, from Eq (19)&(20), it can be easily stated that the perfect tunneling performance of such unit cells will not be as good as that of type-A1. Although The condition for $K_{\parallel}$ will not be satisfied ,such unit cells show a large angular range transparency(it will not show all angle transparency like Type-A1 unit cells even in absence of material loss) (Fig. 4(d)). Moreover, the PTB in DPS region (PSPT) is exemely narrow (also shifted like A-1); see Fig. 4(e) & the difference in field distribution in Fig. 3(e) &5(e). But the notable fact of such arrays is that the band gap properties of type-A2 array are almost identical to type-A1 arrays due to the automatic impedance matching condition (See Eq (18)& the field distributions in Fig 6(c) & (e)). As a result , without satisfying the extra condition with much degrees of freedom, these A2 arrays can be applied in more flexible application purposes.

*Type –B and Type-C unit cells* :
According to Zouhdi et al.[17]: "In terms of Band theory , a wave transmission without modification in phase and amplitude through a PC slab can be achieved if the z directed component of the Bloch vector is equal to zero or, what is just the same, $\cos K_{BZ}(d_1 + d_2) = 1$". This $\cos(K_{BZ}L) = 1$ or band edge condition is automatically fulfilled at subwavelength region for Type A arrays . But for the electrically thick layers, we have found that $K_{BZ} = \pi/L$ is also a similar band edge condition( $\cos(K_{BZ}L) = -1$ ) for achieving perfect tunneling (as in [1]). On the other hand, for PSPT (type-C) $K_{Bz} \neq 0$ or $\pi/L$ [2,3]. As a result, although the magnitude remains same, the phase of the restored wave is modified (Eq 24).So, the only way to achieve wave transmission without modification in phase and amplitude (pure PT) through a PC slab is the band edge condition $\cos(K_{BZ}L) = +1$ or $-1$ with the additional condition $m_{12} = m_{21} = 0$. This condition is also the necessary condition of all angle transparency. Simply, T= **I** is the condition of pure PT where **I** is the identity matrix. This statement is true in general for electrically thick as well as thin layers.

So far we have considered only electrically thin layers. Now we shall discuss electrically thick unit cells. This discussion is also applicable for electrically thin case. For type-B unit cells of Table-1 both electrically thin and thick

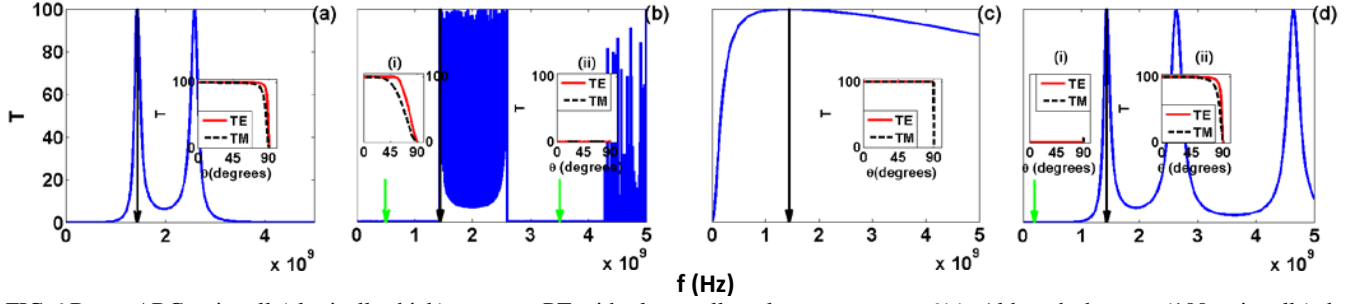

FIG-6.B typeABC unit cell (eletrically thick) supports PT with almost all angle transparency, 6(a). Although the array(100 unit cells) does not support PTB but shows PT at target frequency $f_0$, for normal incidence, 6(b). BGs show CBG performance 6(b). [Prameters: at target frequency $f_0$ = 1.432GHz, $\varepsilon_1$ = 409.42, $\mu_1$ = 1; $\varepsilon_2$ = –1000,$\mu_2$ = 1,$\varepsilon_3$ = 409.42,$\mu_3$ = 1; $d_1 = d_2 = d_3$ = 1mm; Drudedispersive model has been used. Inset of 6(a) is drawn at $f_0$ = 1.432GHz. Inset (i) of 6(b) is at $f_0$ = 1.432GHz. Inset (ii) of 6(b) is for both 0.5GHz & 3.5GHz (both are the centers of corresponding SNG gaps)].

B typeABC unit cell (eletrically thin) supports PT with all angle transparecy, 6(c). The array(100 unit cells) does not support PTB but shows PT at target frequency $f_0$, for normal incidence 6(c). The BG shows CBG performance 6(d) [prameters: at target frequency $f_0$ = 1.432GHz, $\varepsilon_1$ =200.82, $\mu_1$ = 1; $\varepsilon_2$ = –400,$\mu_2$ = 1,$\varepsilon_3$ = 200.82,$\mu_3$ = 1; $d_1 = d_2 = d_3$ = 0.01955mm; Drude dispersive model has been used. Inset of 6(c) is drawn at $f_0$= 1.432GHz. Inset (i) of 6(d) is at 0.2GHz & inset (ii) is at $f_0$ = 1.432GHz (almost all angle transparency is clear).

layers are possible. TheBloch vector should be made zero or $\pi/L$ according to Eq (10) by imposing the condition: $m_{11} = m_{22} = +1$ or $-1$ with the condition $m_{12} = m_{21} = 0$ (for det(T) =1).These conditions are not automatically achieved for the electrically thick layers (and also for thin layers composing of unit cell of the structure DPS SNG DPS).The condition $m_{11} = m_{22}$ can be achieved very easily if we use ABA structure. Interestingly the structure considered by Zhou et al. [1] satisfies the condition $m_{12} = m_{21} = 0$. Unfortunately this interesting fact has been overlooked by the authors in [1]. Instead of $m_{12} = m_{21} = 0$ condition, $m_{12}(f_T) = \eta^2 \cdot m_{21}(f_T)$ condition has been used by the authors (Eq (2) in [1]). Mathematically we can show how, $m_{12} = m_{21} = 0$ condition is automatically achieved when the target PT frequency is the band edge frequency and vice versa. The band edge condition:

$$\frac{1}{2}(m_{11} + m_{22}) = 1 \text{ or } -1 \quad (26)$$

and from the conservation of energy:

$$\det(T) = 1 \text{ or } m_{11}m_{22} - m_{12}m_{21} = 1 \quad (27)$$

we get, $(m_{11} - m_{22}) = (m_{21} - m_{12})$, ignoring absorption, or $-4m_{12}m_{21} = (m_{21} - m_{12})^2$. Now if we consider $m_{11} = m_{22}$, then we find: $m_{12} = m_{21}$. Combining all the equations for $m_{12} = m_{21}$ structure, at the band edge frequency we get:

$$m_{12} = m_{21} = 0 \quad (28)$$

On the other hand, starting from Eq (28) and using Eq (27); the final condition will be found to be nothing but Eq (26). Moreover, using Eq (24) & (25), we can also prove T=**I** and the band edge condition (see Eq (26)) in a much simpler way.However, this condition $m_{12} = m_{21} = 0$ is very important for getting perfect tunneling. Although the structure is electrically thick, it can achieve tranparecy for

Eq (17)). This almost-all-angle transparency for both thepolarizations will not be achieved if $q_{21}= 0$ but $m_{12} \neq 0$ and $m_{21} \neq 0$. Now the important question is: why the electric and magnetic field distributions are quite different in ABAstructure [1] in comparison to Type-A structures (A1 & A2). The answer again comes from the band gap theory. At the band edge, the modes are pseudo propagating modes instead of pure evanscent modes when the structure is DPS|SNG|DPS (Fig 6). According to [17] : "If the impedance has the same value throughout the system, the waves exhibit no reflection. In other words each other (forward and backward propagating waves) do not couple. Thus the phenomena that are connected with reflection, such as the opening of the band gaps, do not appear". In type-B unit cells impedances are not matched. Hence the mathematical formulation for the *E* or *H* field can not be written as two uncoupled independent solutions of the differential equation (2). Due to different impedancevalues, reflected waves should take place and it will cause an opening of BGs (Fig. 6(b)). But at the target PT frequency or band edge (Fig 6(b)), just before (or after)the vanishing of the group velocity (due to the pseudo-propagating and reflected pseudo–propagating modes); decoupling modes exist($K_{BZ}$= 0 or $\pi$). Only for that specific frequencytwo uncoupled independent solutions of *E* or *H* exist. It can be shown using the RHS of the dispersion relation that the field at PT frequency (in ourproposed example, at left band edge frequency of the BG, Fig 6(b)) will be restored after passing through the total array structure. However, no perfect tunneling band will be found for the array of type-B unit cells because just after that specific frequency of perfect tunneling, the modes will start to couple with each other. It will cause destructive interferance and ultimately open the BG. Now, if we consider the band gap property of type-B unit cell array, we may get CBG only for specific cases. But the reason of finding CBG in electrically thick type-B unit cell array is quite different from that of type-A arrays. For our example

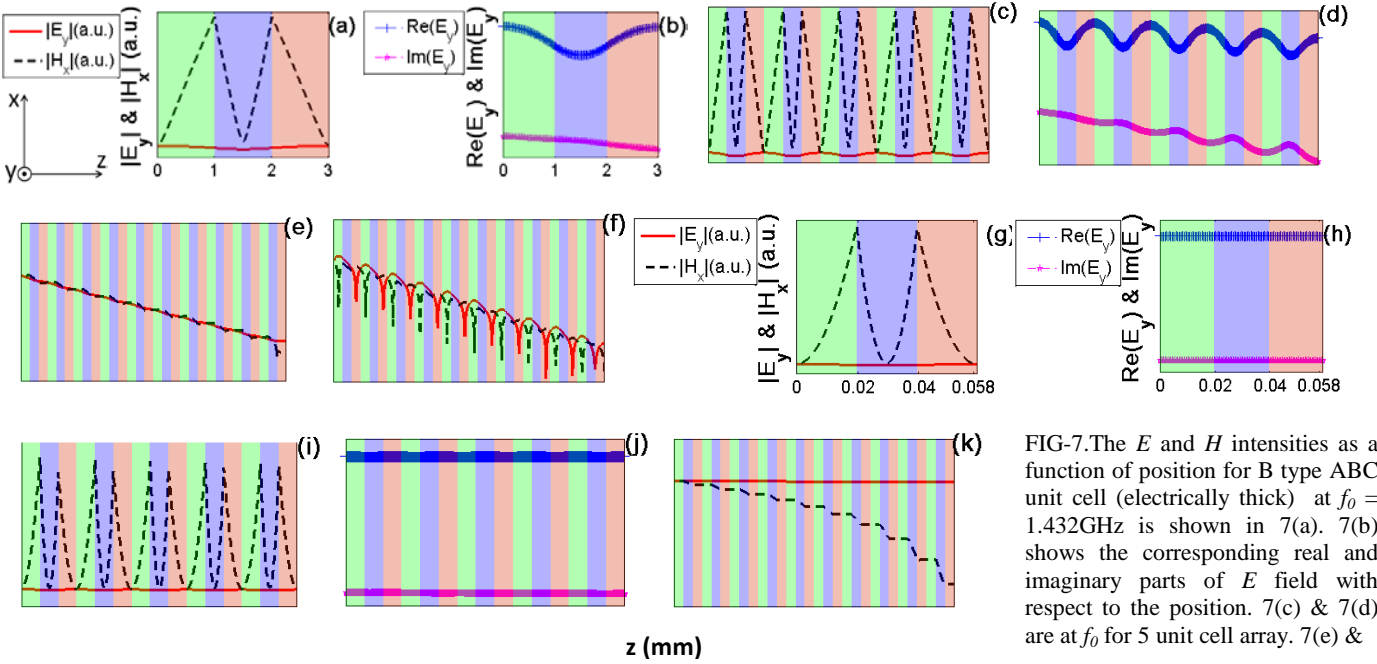

FIG-7. The $E$ and $H$ intensities as a function of position for B type ABC unit cell (electrically thick) at $f_0 = 1.432$GHz is shown in 7(a). 7(b) shows the corresponding real and imaginary parts of $E$ field with respect to the position. 7(c) & 7(d) are at $f_0$ for 5 unit cell array. 7(e) & 7(f) are drawn at mid 0.5GHz & at mid 3.5GHz of the SNG gaps for 10 unit cell array in log scale.

The $E$ and $H$ intensities as a function of position for B type ABC unit cell (electrically thin) at $f_0 = 1.432$GHz is shown in 7(g). 7(h) shows the corresponding real and imaginary parts of $E$ field with respect to the position. 7(i) & 7(j) are at $f_0$ for 5 unit cell array. 7(k) is drawn at 0.20GHz in the SNG gaps for 10 unit cell array in log scale.

(Fig 6(b)), this CBG is found due to the high values of permittivities in all the three layers. Actually $K_{\parallel}$ remains almost the same for any value of the angle of incidence as $|\epsilon \cdot \mu| \gg 1$. On the other hand, the BG equation remains unchanged for both TE & TM modes because of the unchanged value of $F_1$, $F_2$ (see Eq (19)) whether it is TE or TM. But we note that in this case we get real BG which is not possible for type-A arrays. Moreover, this BG due to the pseuodo propagating modes is different from the ZEP gaps [9]. ZEP type of gap is originated due to the interaction of pure evanscent modes which has been mentioned as type-II gap in [20]. Hence it is easily recognisable that although the left band edge frequency is the PT frequency (Fig. 6(b)), it should not be considered similar to 'zero effective phase delay point' [9] like in type-A case.

Another interesting fact is that type-B perfect tunneling is also applicable for electrically thin layers (Fig 6(c) & (d) ). But it is noticable that the conditions of Eq (19) or (20) are not met by them. Again from band theory we can explain this. Actually the band gaps found in DPS|SNG|DPS (actually any combination where complementary material – ENG & MNG have not used together in the set) are not SASN band gaps. PT conditions of Eq (19) and (20) are applicable for unit cell having at least one ENG & MNG layer. Hence although we get almost pure PT at subwavelength limit for type-B electrically thin unit cells, the PT is quite different from the PT reported for type-A cases. Besides, a real BG opens up for Type-B cases. This phenomena can be explained mathematically using Eq(24), (25) & T=**I** condition. We note that in our examples of type-B unit cells, the values of $\beta$ and $\alpha$ are negligibly small (which should be ideally zero). Last but not the least, from Fig. 6 it can be told easily that type-B unit cells may be very good for perfect tunneling (which is experimentally verified in [1]). But from the point of view of BG such thick layers are not inherently as good as type-A arrays because the CBG condition is not achieved automatically.

Finally we shall discuss about type-C unit cell (see [2,3]) and their arrays (electrically thick & thin cases). So far we have shown that the band edge frequency is the perfect tunneling frequency for the unit cells. Besides, we are familiar with the case that when suitable defect layer is introduced to make hetero-structures, the PT frequency is found inside the BG[21, 22] ($K_{BZ}$ is imaginary). But strangely type-C is the only case for which the perfect tunneling frequency is found at the pass band after making array with the unit cells. Based on Eq (24) & (25), we can easily tell that this kind of PT is not pure PT rather it can be considered as phase shifted PT. Type-C unit cells are the most general unit cells satisfying the perfect tunneling condition: $q_{21} = 0$, where no restriction is imposed on the elements of the T matrix (see Eq 16 ). Interestingly, the value of $K_{BZ}$ is real for this type instead of zero or $\pi/L$ or an imaginary value. As T≠**I**, the perfect tunneling performance of type-C unit cells are not good enough like any other types of table-1. More specifically, all angle transparecy can not be achieved in such cells due to the factthat $m_{21} \neq 0$ and $m_{12} \neq 0$. However, phase shifted perfect tunneling will be found at normal incidence for both polar-izations (Fig. 8(a)). We shall consider the structure ofFeng et al. to describe this type-C unit cells [2]. If we makearray with type-C unit cells, at the target perfect tunneling frequency we still get phase shifted PT. But the phase shift due to the array is different from that due the unit cell.

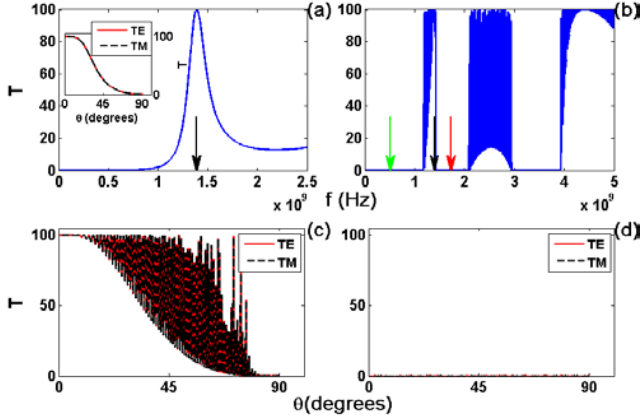
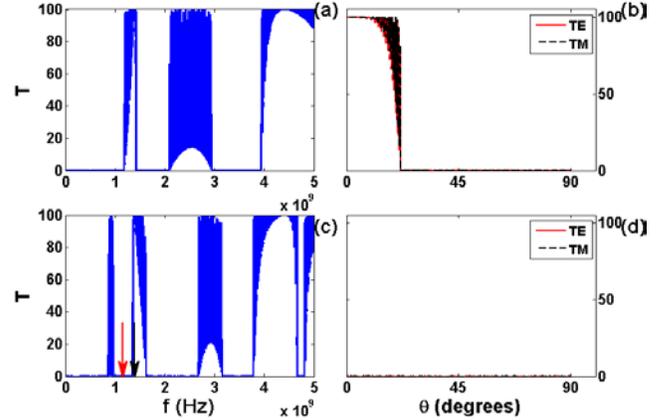

FIG-8. C type ABC unit cell does not support CPT (it shows PT only near the normal incidence) 8(a).[Parameters: at target frequency 1.389 GHz, $\varepsilon_1 = 1$ $\mu_1 = -4.25$, $\varepsilon_2 = 1$ $\mu_2 = 1$, $\varepsilon_3 = -4.24$ $\mu_3 = 1$; $d_1 = d_2 = d_3 = 20$mm; Drude dispersive model has been used. Inset of 8(a) is drawn at the frequency of 1.389GHz (black arrow)].This type possesses extremely narrow PTB when arrayed 8(b).Fig 8(c) is drawn at the target frequency (black arrow).C type ABC type array (100 unit cells) supports CBG fig – 8(d). Both of the gaps are SNG gaps at frequencies of 0.5GHz (Green arrow)& 1.73GHz (Red arrow) respectively. It is notable that a PTBis found in DPS region for array, but its width may not be controllable according to the designer's will.

FIG-9. C type conjugated matched ABCD type array (100 unit cells) showing BG shifting property - 9(a) and 9(c) (See text for detail).[Prameters: at target frequency 1.389 GHz, $\varepsilon_1 = 1$ $\mu_1 = -4.25$, $\varepsilon_2 = 1$ $\mu_2 = 1$, $\varepsilon_3 = -4.24$ $\mu_3 = 1$ with matched layer having $\varepsilon_4 = 2.251$, $\mu_4 = 2.251$; $d_1 = d_2 = d_3 = d_3 = 20$mm; Drude dispersive model has been used. Even after the addition of matched layer, C type cell still shows PT-9(b) at target frequency 1.389GHz (Black arrow). The shifted band gap also shows CBG behavior -9(d), drawn at 1.15GHz (Red arrow) due to imposed conditions.

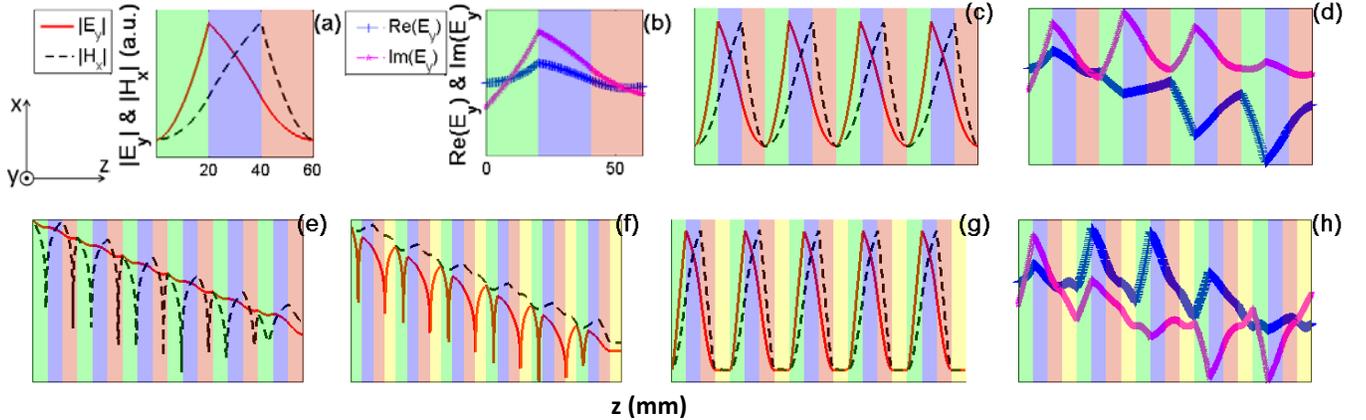

FIG-10. The *E&H* intensities as a function of position for C type ABC unit cell at $f_0 = 1.389$GHz is shown in (a). 10(b) shows the corresponding real and imaginary parts of *E* field with respect to the position. 10(c) & 10(d) are drawn for 4-unit cell array at $f_0 = 1.389$GHz(at pass band), 10(d) shows the corresponding real and imaginary parts of *E* field with respect to the position. 10(e) is drawn for 1.73GHz (at mid of SNG stop band 6 unit cell array, in log scale). 10(f) is at the mid band frequency of 1.15GHz in the shifted BG for the array with 6 ABCD type unit cell, in log scale. 10(g) & 10(h) are drawn for 5-unit cell array of ABCD at target PT frequency 1.389GHz (Pass band) after adding the matched D layer. Field distribution (magnitude only) remains same for ABC & ABCD array at target PT frequency inside the pass band.

Mathematically we can analyze similarly as before for Eq (24) & (25)and hence we can write:

$$\beta_a = \tan^{-1}\left(\frac{1}{\eta_{in} i} \frac{\tilde{m}_{12a}}{\tilde{m}_{11a}}\right) \quad (29)$$

For phase shift in the electric field and similarly for the phase shift in the magnetic field we get:

$$\alpha_a = \tan^{-1}\left(\frac{1}{\eta_{in} i} \frac{\tilde{m}_{21a}}{\tilde{m}_{22a}}\right) \quad (30)$$

Here tilde represents elements of the inverse matrix $(T_a)^{-1}$ the subscript '$a$' refers to the array. The definition of the transfer matrix for the array $T_a$, can be given in a similar way as :

$$T_a = (M_1 \cdot M_2 \cdot M_3)_1 \cdot (M_1 \cdot M_2 \cdot M_3)_2 \ldots (M_1 \cdot M_2 \cdot M_3)_N$$
$$= T_1 \cdot T_2 \ldots T_N = \begin{bmatrix} m_{11a} & m_{12a} \\ m_{21a} & m_{22a} \end{bmatrix} \quad (31)$$

However, decoupling mode theory reported in [17] is not applicable to this PSPT phenomenon because there must be reflected waves in each of the layers at PSPT frequency for the type-C array structures. It is noticable that a wide PSPT band is found in the DPS region when the layers are electrically very thick (Fig 9(a) & (d)). Similar stable PSPT band has been observed in type-A array structure. Now, if we consider the band gap property of type-C unit cell array, we may get CBG only for specific cases. But the reason of finding CBG with electrically thick type-C unit cell array is quite different from that in the type-A & B arrays. In our example (Fig 8(d)), this CBG is found due to the imposed condition of CBG reported in [10, 23]. Our structure is ENG|MNG|DPS type and detailed physical explanation of CBG formation can be found in [23].But this BG is similar to the type-II gap reported in [20] ( due to interaction of evascent waves, see Fig 11 (c)).

Another important fact is that type-C arrays can support a new phenomenonthat we can call 'band gap shifting' in which we add an extra matched layer D with the previous perfect tunneling ABC unit cell of [2] to make $\sum \epsilon_i d_i = \sum \mu_i d_i = 0$ at PT frequency. Making an array of the form $(ABCD)^n$, we find that the BG at the RHS of the PT frequency in the array of ABC jumps to the LHS after adding the matched layer D(Fig 9(a) & (c)). The new BG will also show CBG (Fig 9(d)) just like the CBG in the previous RHS BG of $(ABC)^n$ structure. However, the field distribution is differenet for the new LHS BG of the $(ABCD)^n$ (see Fig. 10(c)& 10(g)). The BG is still formed due to the interaction of evanscent modes (similar to type-II gap in [20]).It is interesting that even after the addition of an extra layer, the PSPT frequency of the ABCD layer remains thesame as that of the ABC layer (Fig 9 (c)). Besides, the field distribution (magnitude) is not affected due to the addition of the extra matched layer (Fig 10(g)).ThePT band in the DPS region will be a little bit affected and reduced for the array of ABCD (Fig (a)& (d)). Feng et al. [2] have experimentally checked the PT performance of their unit cell. Detail physical and mathematical explanation of this interesting phenomena- 'Band Gap Shifting', will be discussed elesewhere in future. For electrically thin type-C unit cells, we expect similar results. Finally we can conclude that although the PT performance is not good enough for the type-C unit cells, their array performance is good enough due to simultanous PTB and CBG even with elecrtically thick layers. Last but not least, although in [2,3] PSPT has been achieved with metamaterials,PSPT is also achievable with unit cells made up of only DPS layers.

**Conclusion:**
We have proposed a compact calssification of unit cells satisfying the condition $q_{21} = 0$ and showing PTmade up of SNG metamaterialsbased on our derived specific relations between Q & T matrices. The physical explanation of PT in the arrayshas been made based on band theory& under the assumption of no loss. As in the microwave range, the effect of loss is not too high. We believe are proposed structures are very promising for future applicationse.g. for optical fibers, subwavelength imaging, broadband data transmission, anetnna engineering [11-13, 24-29] and so on.We predict our electromagnetic ideas can also be applied on semiconductor structures for perfect tunneling and new types of artificial electronic band gaps.

[1] L. Zhou, W. Wen, C. T. Chan, and P. Sheng, Phys. Rev. Lett. 94, 243905 (2005).
[2] T. Feng *et al*, Phys. Rev. E. 79, 026601 (2009).
[3] G. Castaldi, I. Gallina, V. Galdi, A. Alù, N. Engheta, Phys. Rev. B 83, 081105(R) (2011).
[4] E. Cojocaru, Progress In Electromagnetics Research, Vol. 113, 227-249, 2011.
[5] A. Alu, N. Engheta, Antennas and Propagation, IEEE Transactions on (Volume: 51, Issue: 10)
[6] L. Jelinek, J. D Baena, J. Voves, R. Marques New Journal of Physics, Vol. 13, 2011.
[7] M. G. Silveirinha, N. Engheta Phys. Rev. Lett. 110, 213902 (2013).
[8] J. Li, L. Zhou, C. T. Chan,P. Sheng, Phys. Rev. Lett. 90, 083901 (2003).
[9] H. Jiang*et al*,, Phys. Rev. E 69, 066607 (2004).
[10] I. V. Shadrivov, A. A. Sukhorukov, Y. S. Kivshar Phys. Rev. Lett. 95, 193903 (2005).
[11] M. M. Hasan, D. S. Kumar, M. R. C. Mahdy, D. N. Hasan, M. A. Matin, "Robust Optical Fiber Using Single Negative Metamaterial Cladding" ,IEEE Photonics Technology Letters, Vol. 25, pp. 1043-1046 , June 2013.
[12] E. J. Smith , Z. Liu, Y. Mei, O. G. Schmidt, Nano Lett., 2010, *10* (1), pp 1–5, DOI: 10.1021/nl900550j.

[13] D.J. Hu, G Alagappan, Y.K. Yeo, P.P. Shum, P. Wu, Optics Express, Vol. 18, 2010.
[14] S. B. Cavalcanti, M. de Dios-Leyva, E. Reyes-Gómez, L. E. Oliveira, Phys. Rev. B 74, 153102 (2006).
[15] S. J. Orfanidis, *Electromagnetic Wave and Antennas*, (Piscataway, New Jersey, 2008)
[16] P. Yeh, A. Yariv, C.S. Hong, JOSA, Vol. 67, Issue 4, pp. 423-438 (1977)
[17] S. Zouhdi, A. V. Dorofeenko, A. M. Merzlikin, A. P. Vinogradov, Phys. Rev. B 75, 035125 (2007)
[18] Y Weng, ZG Wang, H Chen, Physical Rev. E 75, 046601 (2007)
[19] Y. Xiang, X. Dai, S. Wen, Z. Tang, D. Fan, Applied Physics B, Vol. 103, 2011.
[20] L.G. Wang, H. Chan, S.Y. Zhu, Phys. Rev. B 70, 245102 (2004).
[21] G. Guan *et al*, Appl. Phys. Lett. 88, 211112 (2006)
[22] Y. Chen, J. Opt. Soc. Am. B, Vol. 25, 2008.
[23] T.B. Wang, J.W. Dong, C.P. Yin, H.Z. WangPhysics Letters A, Vol. 373, 2008.
[24] M. R. C. Mahdy, M. R. A Zuboraj, A. A.N. Ovi, and M. A. Matin, "'A Novel Design Algorithm' And Practical Realization of Rectangular atch Antenna Loaded With SNG Metamaterial," *Progress In Electromagnetics Research M*, Vol. 17, 13-27, 201.
[25] M.R.C.Mahdy, M.R.A. Zuboraj, A. Al Noman Ovi, and M. A.Matin, "An Idea of Additional Modified Modes in Rectangular Patch Antennas Loaded With Metamaterial" IEEE Antennas and Wireless Propagation Letters, vol. 10, pp. 869-872, August 2011.
[26] SaimoomFerdous, Md.AbabilHossain, Shah Md Mahmud HasanChowdhury, MahdyRahmanChowdhuryMahdy and Md. Abdul Matin, "Reduced and Conventional size Multi-Band Circular Patch Antennas Loaded with Metamaterials", IET Microwaves , Antennas and Propagation , vol. 7 , pp. 768-776 , July 2013.
[27] Mehedi Hassan, M. R. C. Mahdy, Gazi M.Hasan, LutfaAkter, "A novel miniaturized triple-band antenna" 7th International Conference on Electrical & Computer Engineering (ICECE), Dhaka, Bangladesh, pp. 702-705 , Dec 2012.
[28] M. R. C. Mahdy, M. R. A. Zuboraj, A. A. N. Ovi and M. A. Matin "Novel Design of Triple Band Rectangular Patch Antenna Loaded with Metamaterial" Progress In Electromagnetics Research Letters, Vol. 21, 99-107, 2011.
[29] Shah Mahmud HasanChowdhury, Md. AbabilHossain, Md. SaimoomFerdous, MahdyRahmanChowdhuryMahdy, Md. Abdul Matin, "Conceptual and practical realization of reduced size multi-band circular microstrip patch antenna loaded with MNG metamaterial" 7th International Conference on Electrical & Computer Engineering (ICECE), Dhaka, Bangladesh, pp. 834-837 , Dec 2012.